\documentclass[a4paper,onecolumn,conference]{IEEEtran}
\usepackage[T1]{fontenc}
\usepackage[latin9]{inputenc}
\usepackage{textcomp}
\usepackage{subfig}
\usepackage{amsmath}
\usepackage{cite}
\usepackage[ruled,vlined]{algorithm2e}
\usepackage{bbm}
\usepackage{graphicx}
\usepackage{epstopdf}
\usepackage{balance}
\usepackage{amssymb}
\usepackage{array}
\usepackage{fancyhdr}
\begin{document}

\title{Efficient Data Compression with Error Bound \\ Guarantee in Wireless Sensor Networks}

\author{\IEEEauthorblockN{Mohammad Abu Alsheikh\IEEEauthorrefmark{1}\IEEEauthorrefmark{2},
Puay Kai Poh\IEEEauthorrefmark{3},
Shaowei Lin\IEEEauthorrefmark{2},
Hwee-Pink Tan\IEEEauthorrefmark{2},
Dusit Niyato\IEEEauthorrefmark{1}}

\IEEEauthorblockA{\IEEEauthorrefmark{1}School of Computer Engineering, Nanyang Technological University, Singapore 639798}
\IEEEauthorblockA{\IEEEauthorrefmark{2}Sense and Sense-abilities Programme, Institute for Infocomm Research, Singapore 138632}
\IEEEauthorblockA{\IEEEauthorrefmark{3}National University of Singapore, Singapore 138632}

}

\maketitle
\begin{abstract}
\textbf{We present a data compression and dimensionality reduction scheme for data fusion and aggregation applications to prevent data congestion and reduce energy consumption at network connecting points such as cluster heads and gateways. Our in-network approach can be easily tuned to analyze the data temporal or spatial correlation using an unsupervised neural network scheme, namely the autoencoders. In particular, our algorithm extracts intrinsic data features from previously collected historical samples to transform the raw data into a low dimensional representation. Moreover, the proposed framework provides an error bound guarantee mechanism. We evaluate the proposed solution using real-world data sets and compare it with traditional methods for temporal and spatial data compression. The experimental validation reveals that our approach outperforms several existing wireless sensor network's data compression methods in terms of compression efficiency and signal reconstruction.} 
\end{abstract}


\section{Introduction}

Many wireless sensor networks today play an important role in collecting big amounts of real-time sensing data over large areas. A gateway, for instance, may gather data from the sensor network before sending it over long distances to a base station. The sensor network might also have cluster heads that aggregate the data from its corresponding nodes for transport to other cluster heads. Data compression and dimensionality reduction in wireless sensor networks (WSNs) refer to the problem of encoding the data collected from sensor nodes using fewer bits. Compression at cluster heads, gateways, or even within a sensor node with multiple sensing units, is one key ingredient in prolonging network lifetime \cite{fasolo2007network}. Moreover, archiving the collected data for several years requires a tremendous capacity of storage that ranges from terabytes to petabytes \cite{gandhi2009gamps}. However, traditional data compression schemes from information and coding theory cannot be directly applied to a resource limited framework like WSNs as they are designed to optimize storage rather than energy consumption \cite{razzaque2013compression}. 

Lossy compression methods in WSNs are preferable over the lossless ones as they provide better compression ratio at lower computational cost \cite{razzaque2013compression}. However, most traditional lossy data compression algorithms in WSNs lack an error guarantee mechanism due to the high computational demand of data decompression and reconstruction \cite{razzaque2013compression}. Therefore, many existing lossy methods rely on statistical analysis to examine the probability of data loss or assume the data loss is due to noise effects such that the loss can be ignored \cite{Srisooksai201237}. Moreover, the complexity of the decompression routine becomes critical when the data destination is another node in the network. Thus, the computational complexity of data decompression is still an important concern.

The above discussion motivates the need for one solution that collectively supports the aforementioned design essentials. Briefly, our main contributions are as follows.
\begin{itemize}
	\item We propose a low-cost (both compression and decompression) lossy compression technique with error bound guarantee. The routines for compression and decompression are implemented using only linear and sigmoidal operations.
	\item Unlike many traditional methods, our method is easily customized for both temporal and spatial compression. This allows the design of a uniform sensing framework that does not require many dedicated compression solutions, one for each application.
\item Experiments over real world data sets show that the algorithm outperforms several well-known and traditional methods for data compression in WSNs.
\end{itemize}

\section{Related works}\label{sec:related}

In this section, we identify a variety of coding schemes in the literature \cite{Srisooksai201237,zordan2012compress,razzaque2013compression}, and discuss some important considerations for signal compression in WSNs.

The lightweight temporal compression (LTC) algorithm \cite{schoellhammer2004lightweight} is an efficient method that finds a piece-wise linear representation for time series in sensor data. Unfortunately, it performs poorly if the sensor readings fluctuate frequently, even when the fluctuations follow some fixed patterns over time. Moreover, as its name implies, it can only be used for temporal data compression. Principal component analysis (PCA) has been widely used to extract dominant linear features in sensor readings \cite{rooshenas2010reducing}. Another large class of lossy data compression techniques involves the transformation of the raw data into other data domains. Examples of these methods are based on discrete Fourier transforms (DFT) and fast Fourier transforms (FFT) \cite{zordan2012compress} and the different types of discrete cosine transforms (DCT) \cite{quer2009interplay}. However, such algorithms suffer from poor performance when used to compress data spatially or when noise is present in the collected readings.

If a sparse representation for the given signals is known, compressive sensing (CS) is another framework for transforming the signal into an efficient compressed form, which will be used later to recover an approximation of the original signal, e.g., \cite{xiang2011compressed}. However, the assumption of sparsity in the input signal can be highly restrictive, as the sensor data may not be sparse in the time domain, the frequency domain, or even in some other traditional domains. Moreover, introducing a few noisy readings may corrupt the sparse data representation, and the complexity of CS's data decoding hinders the development of an error bound for such lossy methods. For dictionary-based lossless data compression in WSNs, the Sensor Lempel-Ziv-Welch (S-LZW) algorithm \cite{sadler2006data} is a typical approach. However, S-LZW does not consider the temporal and spatial characteristics of collected data which, if used, can significantly enhance the compression performance.

\section{Neural Autoencoders}\label{sec:preliminaries}

\begin{figure}
\begin{centering}
\includegraphics[width=0.45\columnwidth]{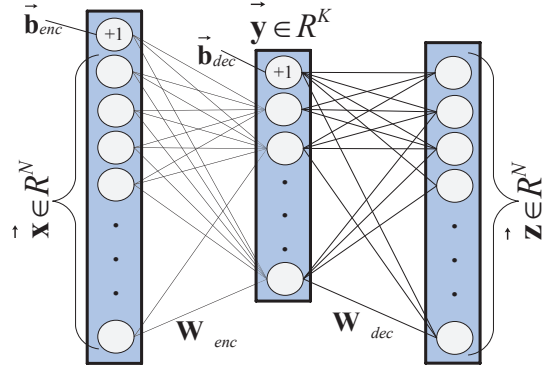}
\par\end{centering}
\caption{\label{fig:basic_autoencoder}Using AE to project the data to a lower dimensional representation ($K\ll N$).}
\end{figure}

An autoencoder (or auto-associative neural network encoder) is a three-layer neural network that maps an input vector $\vec{\mathbf{x}}\in\mathbb{R}^{N}$ to a hidden representation $\vec{\mathbf{y}}\in\mathbb{R}^{K}$ and finally to an output vector $\vec{\mathbf{z}} \in\mathbb{R}^{N}$ that approximates the input $\vec{\mathbf{x}}$, as shown in Figure~\ref{fig:basic_autoencoder}. The vectors satisfy
\begin{subequations}
\begin{equation}
\vec{\mathbf{y}}=F\left(\mathbf{W}_{enc}\vec{\mathbf{x}}+\vec{\mathbf{b}}_{enc}\right)
\end{equation}
\begin{equation}
\vec{\mathbf{z}}_{\boldsymbol{\theta} }(\vec{\mathbf{x}})=F\left(\mathbf{W}_{dec}\vec{\mathbf{y}}+\vec{\mathbf{b}}_{dec}\right)
\end{equation}
\begin{equation}
F\left(\upsilon\right)=\frac{1}{1+\exp(-\upsilon)}
\end{equation}
\end{subequations}
where $\boldsymbol{\theta} :=[\mathbf{W}_{enc},\vec{\mathbf{b}}_{enc},\mathbf{W}_{dec},\vec{\mathbf{b}}_{dec}]$ are real-valued parameters that must be learned by a suitable training algorithm, and $F\left(\cdot\right)$ is the sigmoidal logistic function (other nonlinear function such as the hyperbolic tangent can also be used). The parameters $\mathbf{W}_{enc}$ and $\vec{\mathbf{b}}_{enc}$ are the encoding weight matrix and bias respectively, while $\mathbf{W}_{dec}$ and $\vec{\mathbf{b}}_{dec}$ are the decoding weight matrix and bias. The entries of $\vec{\mathbf{y}}$ and $\vec{\mathbf{z}}$ are sometimes called activations.

To learn optimal neural weights $\boldsymbol{\theta}$ using training data $\mathbf{D}$, we define the cost function of the basic autoencoder (AE):
\begin{equation}\label{eq:cost_function}
\begin{aligned}\Gamma_{\text{AE}}\left(\boldsymbol{\theta},\mathbf{D}\right) =\frac{1}{\vert\mathbf{D}\vert }\sum_{\vec{\mathbf{x}}\in\mathbf{D}}{\frac{1}{2}\left\Vert \vec{\mathbf{x}}-\vec{\mathbf{z}}_{\boldsymbol{\theta} }(\vec{\mathbf{x}})\right\Vert ^{2}}.
\end{aligned}
\end{equation}
This function penalizes the difference between each input data vector $\vec{\mathbf{x}}$ and its reconstruction $\vec{\mathbf{z}}_{\boldsymbol{\theta} }(\vec{\mathbf{x}})$. Consequently, the optimal neural weights may be computed using standard optimization algorithms such as the limited memory Broyden\textendash{}Fletcher\textendash{}Goldfarb\textendash{}Shanno (L-BFGS) algorithm.

Different variants of the basic AE have been introduced in the literature to discourage the neural network from overfitting the training data. Generally speaking, these regularization methods penalize the neural weight characteristics~or the hidden layer sparsity characteristics.

\textbf{Weight decaying autoencoder (WAE)}: In this variant, the cost function is defined with an extra weight decay term:
\begin{equation}
\Gamma_{\text{WAE}}\left(\boldsymbol{\theta},\mathbf{D}\right)= \Gamma_{\text{AE}}\left(\boldsymbol{\theta},\mathbf{D}\right)+
\frac{\beta}{2}\left( \left\Vert \mathbf{W}_{enc} \right\Vert^2 +  \left\Vert \mathbf{W}_{dec} \right\Vert^2 \right)
%\frac{\beta}{2}\left(\sum_{i=1}^{N}\sum_{j=1}^{K}\left(w_{ij}^{(enc)}\right)^{2}+\sum_{l=1}^{K}\sum_{o=1}^{N}\left(w_{lo}^{(dec)}\right)^{2}\right)
\end{equation}
where $\left\Vert \mathbf{W} \right\Vert^2$ represents the sum of the squares of the entries of a matrix $\mathbf{W}$, and $\beta$ is a hyperparameter%
\footnote{A hyperparameter is a variable that is selected a priori. This differentiates
it from a model parameter, e.g., the encoding weight, which is adjusted during the learning process.%
} 
that controls the contribution from the weight decay term.

\textbf{Sparse autoencoder (SAE)}: This version extracts a sparse data representation at the hidden layer, i.e. we want most of the entries of $\vec{\mathbf{y}}$ to be close to zero. Sparsity is encouraged by adding the Kullback\textendash{}Leibler (KL) divergence function \cite{ng2011sparse}:
\begin{subequations}
\begin{equation}
\Gamma_{\text{SAE}}\left(\boldsymbol{\theta},\mathbf{D}\right)=\Gamma_{\text{WAE}}\left(\boldsymbol{\theta},\mathbf{D}\right)+\eta\sum_{k=1}^{K}\text{KL}(\rho||\hat{\rho_{k}})
\end{equation}
\begin{equation}
\text{KL}(\rho||\hat{\rho_{k}})=\rho\log_{e}\frac{\rho}{\hat{\rho_{k}}}+\left(1-\rho\right)\log_{e}\left(\frac{1-\rho}{1-\hat{\rho_{k}}}\right)
\end{equation}
\end{subequations}
where $\eta$ is a hyperparameter that controls the sparsity weight, $\rho$ is the sparsity parameter (target activation) that is chosen to be close to zero, and $\hat{\rho_{k}}$ is the average activation of the $k$-th node in the hidden layer.

\vspace{-2mm}
\section{Lossy compression with error bound guarantee}\label{sec:framework}

\begin{figure}
\begin{centering}
\includegraphics[width=0.45\columnwidth]{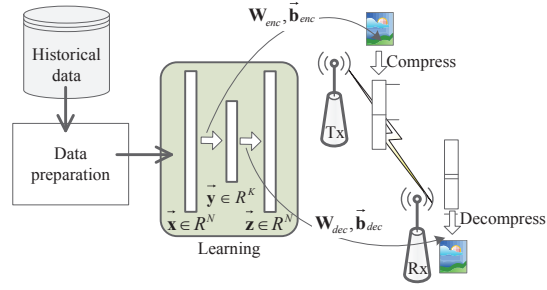}
\par\end{centering}

\caption{\label{fig:autoencoder_adoption}AE adoption for data compression and dimensionality reduction in wireless sensor networks. Initially, the network's parameters $\mathbf{W}_{enc},\vec{\mathbf{b}}_{enc},\mathbf{W}_{dec},$ and $\vec{\mathbf{b}}_{dec}$ are adjusted during the learning stage (offline mode). Subsequently, the encoding part will be executed in the transmitter side (Tx) to achieve a compressed representation of the data. Then the receiver (Rx) will deploy the decoding part to recover a proper approximation of the original signal.}
\end{figure}

We propose to apply the autoencoder to the data compression and dimensionality reduction problem in WSNs to represent the captured data using fewer bits as demonstrated in Figure \ref{fig:autoencoder_adoption}. This is motivated by reasons related to WSN characteristics, as well as the ability of AEs to automatically extract features in the data. Firstly, similar to other lossy data compression algorithms, it is important to realize that AEs are used to extract a suitable, low-dimensional, code representation that retains most of the information content of the original data. This process of automatic feature extraction is not, by any means, intended to randomly discard data items, but instead to find better data representation domains. Secondly, sensor networks are used to collect data in a variety of distinct situations each with its network structure and data patterns. Therefore, the designer must be familiar with a collection of temporal and spatial compression algorithms to support each case. In contrast, the proposed algorithm has the flexibility of supporting many scenarios using one technique. Thirdly, AEs are commonly used to extract intrinsic features that can be used by several data analysis, manipulations, storage, communications, and visualization algorithms \cite{hinton2006reducing}. Further, AEs with nonlinear activation transfer functions, such as the logistic regression, can learn more representative features than the well-known PCA algorithm \cite{japkowicz2000nonlinear}. Fourthly, the distributed data compression alleviates the need for data archiving and storage solutions (for such lossy data archiving solution on database systems, please see \cite{gandhi2009gamps}).
Indeed, the centralized solutions focus on data compression and archiving into the database systems, without considering the bandwidth and the energy limitations during the data funneling and aggregation. Finally, after learning the AE's parameters, the process of data encoding and decoding can be simply programmed with a few lines of code. On the one hand, the simplicity of the encoding process is important as the nodes are resource limited devices. On the other hand, the decoder complexity is crucial when sending data between the sensors or when dealing with thousands of sensor nodes sending their compressed
data continually to a central base station, i.e., as the base station will be required to decompress big data set.

\subsection{Error bound mechanism}
In some applications, it is important to provide a guarantee that the reconstructed signal is close to the original (source). The error bound $\epsilon_{bound}$ is defined as the maximum acceptable difference between each collected reading by the sensor and the recovered one by the receiver after receiving the compressed representation. Basically, the error bound is tuned by considering several factors such as the application requirements and the used sensors' precision. For example, the RM Young/05103 wind monitor sensor \cite{rm_young05103} measures the wind speed and direction with accuracy of 0.3 m/s and $5^{\circ}\text{C}$, respectively. Thus, setting
the error bound to be equal to the sensor accuracy may be an acceptable design basis.

\begin{figure}
\begin{centering}
\includegraphics[width=0.45\columnwidth]{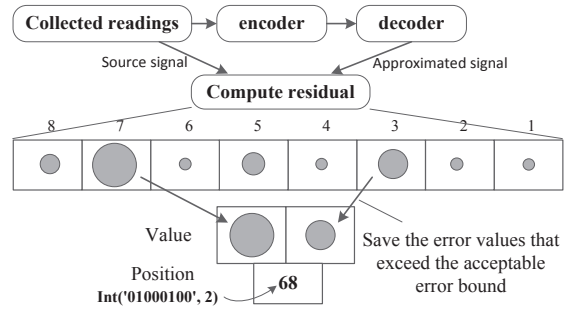}
\par\end{centering}

\caption{\label{fig:error_bound}The error bound mechanism performed by the transmitting node.}
\end{figure}

Our method first computes the residual $\vec{\mathbf{r}} = \vec{\mathbf{p}}-\vec{\mathbf{q}}$ between the source $\vec{\mathbf{p}}$ and the recovered data $\vec{\mathbf{q}}$, as shown in Figure \ref{fig:error_bound}. Any entry of the residual vector exceeding the bound $\epsilon_{bound}$ will be transmitted, using the residual code
\begin{equation}
\vec{\boldsymbol{\varepsilon}}=\text{residualCode}(\vec{\mathbf{r}}, \epsilon_{bound})=\left(\mathbbm{1}_{I}, \left(r_{i}\right)_{i\in I}\right)
\end{equation}
where $I \subset \{1, \ldots, N\}$ is the set of indices $i$ where $r_{i}>\epsilon_{bound}$ and $\mathbbm{1}_{I}$ is the indicator vector for the subset $I$, i.e. $(\mathbbm{1}_{I})_i = 1$ if $i \in I$ and $(\mathbbm{1}_{I})_i = 0$ if $i \notin I$. Conversely, given the code $\vec{\boldsymbol{\varepsilon}}$, it is easy to compute an estimate of the original residual by constructing a vector whose zeros are determined by $\mathbbm{1}_{I}$ and whose non-zero entries are given by $\left(r_{i}\right)_{i\in I}$. We denote~this vector as $\mbox{residual}(\vec{\boldsymbol{\varepsilon}})$.

\subsection{Data sphering}
The entries of the output vector $\vec{\mathbf{z}}$ of the AE come from the sigmoid function, so they are all between $0$ and $1$. Because the AE attempts to reconstruct the input vector $\vec{\mathbf{x}} \in \mathbb{R}^N$, we need to normalize our input data so that the entries are also between $0$ and $1$. Moreover, for the AE to work, the input data vectors must be distributed somewhat uniformly near the unit sphere in $\mathbb{R}^{N}$. This process is called data sphering \cite{ng2011sparse}. One simple method involves truncating readings that lie outside three standard deviations from the vector mean, and rescaling the remaining readings so that they lie between $0.1$ and $0.9$. In particular, the formula is
\begin{equation} \label{eq:truncation}
\begin{split}
\vec{\mathbf{x}} & = \text{normalize}(\vec{\mathbf{p}},\sigma) \\
	&= 0.5 + \frac{0.4}{3\sigma} \max\left(\min\left(\vec{\mathbf{p}}-\text{mean}(\vec{\mathbf{p}}),3\sigma\right),-3\sigma\right)
\end{split}
\end{equation}
where $\vec{\mathbf{p}}$ is the data vector and $\sigma$ is the standard deviation of the entries of $\vec{\mathbf{p}}-\text{mean}(\vec{\mathbf{p}})$ over all $\vec{\mathbf{p}}$ in the training data set. Furthermore, assuming the data is normally distributed, the probability that a reading is located within three standard deviations from the mean is $99.7\%$. Conversely, given the mean $m$, the original data vector $\vec{\mathbf{p}}$ may be reconstructed (up to truncated outliers) using the formula:
\begin{equation} \label{eq:detruncation}
\begin{split}
\text{denormalize}(\vec{\mathbf{x}},m,\sigma) = \frac{3\sigma}{0.4} (\vec{\mathbf{x}}-0.5)+m.
\end{split}
\end{equation}

\subsection{Training, encoding and decoding}
After describing different components of our algorithms, we are now ready to put them together. We assume that all the data mentioned in this section have been aligned and that missing values have been filled in. For the training data $\mathbf{D}$, we also ensure that outliers were removed and that readings were normalized. Let $\sigma$ denote the standard deviation used in the normalization of the data.

We first learn optimal weights $\boldsymbol{\theta}$ for the autoencoder by minimizing the cost function $\Gamma_{\text{WAE}}(\boldsymbol{\theta},\mathbf{D})$ using the L-BFGS algorithm. This computationally-intensive process only occurs once at the start of our network deployment, and the parameters $\boldsymbol{\theta}, \sigma$ are distributed to the transmitters and receivers.

The algorithms for compressing and decompressing the sensor readings are outlined in Algorithms \ref{alg:compression} and \ref{alg:decompression} respectively. For our experiments, we send the compressed signal $(\vec{\mathbf{y}},\vec{\boldsymbol{\varepsilon}},m)$ using floating point representation for the real numbers and binary string for the indicator vector $\mathbbm{1}_{I}$ in $\vec{\boldsymbol{\varepsilon}}$. Note that all the steps have low computational complexity. Here, we also see why decoder complexity in algorithms like compressed sensing impedes the provision of error bound guarantees.

\begin{algorithm}
	\DontPrintSemicolon
	\KwIn{readings $\vec{\mathbf{p}}$; parameters $\sigma, \mathbf{W}_{enc},\vec{\mathbf{b}}_{enc},\mathbf{W}_{dec},\vec{\mathbf{b}}_{dec}$}
	\KwOut{signal $\vec{\mathbf{y}},\vec{\boldsymbol{\varepsilon}},m$}
	\Begin{
	$m \leftarrow \text{mean}(\vec{\mathbf{p}})$\;
	$\vec{\mathbf{x}} \leftarrow \text{normalize}(\vec{\mathbf{p}},\sigma)$\;
	$\vec{\mathbf{y}} \leftarrow F (\mathbf{W}_{enc}\vec{\mathbf{x}}+\vec{\mathbf{b}}_{enc})$\;
	$\vec{\mathbf{z}} \leftarrow F(\mathbf{W}_{dec}\vec{\mathbf{y}}+\vec{\mathbf{b}}_{dec})$\;
	$\vec{\mathbf{q}} \leftarrow \text{denormalize}(\vec{\mathbf{z}},m,\sigma)$\;
	$\vec{\boldsymbol{\varepsilon}} \leftarrow  \text{residualCode}(\vec{\mathbf{p}}-\vec{\mathbf{q}}, \epsilon_{bound})$\;
	}
\caption{The online data compression \label{alg:compression}}
\end{algorithm}

\begin{algorithm}
	\DontPrintSemicolon
	\KwIn{signal $\vec{\mathbf{y}},\vec{\boldsymbol{\varepsilon}},m$; parameters $\sigma,\mathbf{W}_{dec},\vec{\mathbf{b}}_{dec}$}
	\KwOut{reconstruction $\vec{\mathbf{p}}$}
	\Begin{
	$\vec{\mathbf{z}} \leftarrow F(\mathbf{W}_{dec}\vec{\mathbf{y}}+\vec{\mathbf{b}}_{dec})$\;
	$\vec{\mathbf{q}} \leftarrow \text{denormalize}(\vec{\mathbf{z}},m,\sigma)$\;
        $\vec{\mathbf{r}} \leftarrow \text{residual} (\vec{\boldsymbol{\varepsilon}})$\;
	$\vec{\mathbf{p}} \leftarrow \vec{\mathbf{q}} + \vec{\mathbf{r}} $\;
	}
\caption{The online data decompression \label{alg:decompression}}
\end{algorithm}

\vfill\eject
\section{Experimental results }\label{sec:experimental}

We evaluate the performance of the proposed algorithm using data from actual sensor test beds. Our data set is divided into 10 random folds for training and testing. In each experiment, the system is trained using 9 folds and tested using the last fold. Due to randomness in initializing the neural weights, we conduct each experiment 20 times to ensure consistency in the test results. Therefore, the system performance presented is the average obtained from 200 experiments. Our implementation adopts the L-BFGS algorithm \cite{byrd1995limited} to tune the AE's weights during the learning stage. We define the following error metrics:
\begin{subequations}
\begin{equation}
\text{Mean absolute error}=\epsilon_{abs}=\frac{1}{N}\sum_{i=1}^{N}\left|p_{i}-q_{i}\right|
\end{equation}
\begin{equation}
\text{Relative error}=\epsilon_{rel}=\frac{\sum_{i=1}^{N}\left|p_{i}-q_{i}\right|^{2}}{\sum_{i=1}^{N}p_{i}^{2}}\times 100
\end{equation}
\end{subequations}
where $p_{i}$ is the $i-$th entry of the input vector $\vec{\mathbf{p}} \in \mathbb{R}^N$ and $q_{i}$ is the reconstructed value for $p_{i}$. To measure the extent that the data is being compressed, we use the following metric:
\begin{equation}
\text{Compression ratio}=CR=\left(1-\frac{B(\vec{\mathbf{y}})+ B(\vec{\boldsymbol{\varepsilon}})}{B(\vec{\mathbf{p}})}\right)\times100
\end{equation}
where $B(\vec{\mathbf{y}})$, $B(\vec{\boldsymbol{\varepsilon}})$, and $B(\vec{\mathbf{p}})$ are the number of bits used to represent the compressed, the residual, and the original data, respectively. We evaluate our solution using meteorological data set from the Grand-St-Bernard deployment \cite{Sensorscope_Grand}. We use data from 23 sensors that collect surface temperature readings between Switzerland and Italy at an elevation of 2.3km. This data set contains readings ranging from $-32^{\circ}\text{C}$ to $48^{\circ}\text{C}$, though observations suggest that the maximum and minimum values are most likely from a malfunctioning sensor node.

\begin{figure*}
\begin{centering}
\subfloat[{\label{fig:experimental_validation_stbernard_1}\small{}Spatial compression: Compression ratio vs. relative error $\epsilon_{rel}$.}]{\begin{centering}
\includegraphics[width=0.33\columnwidth]{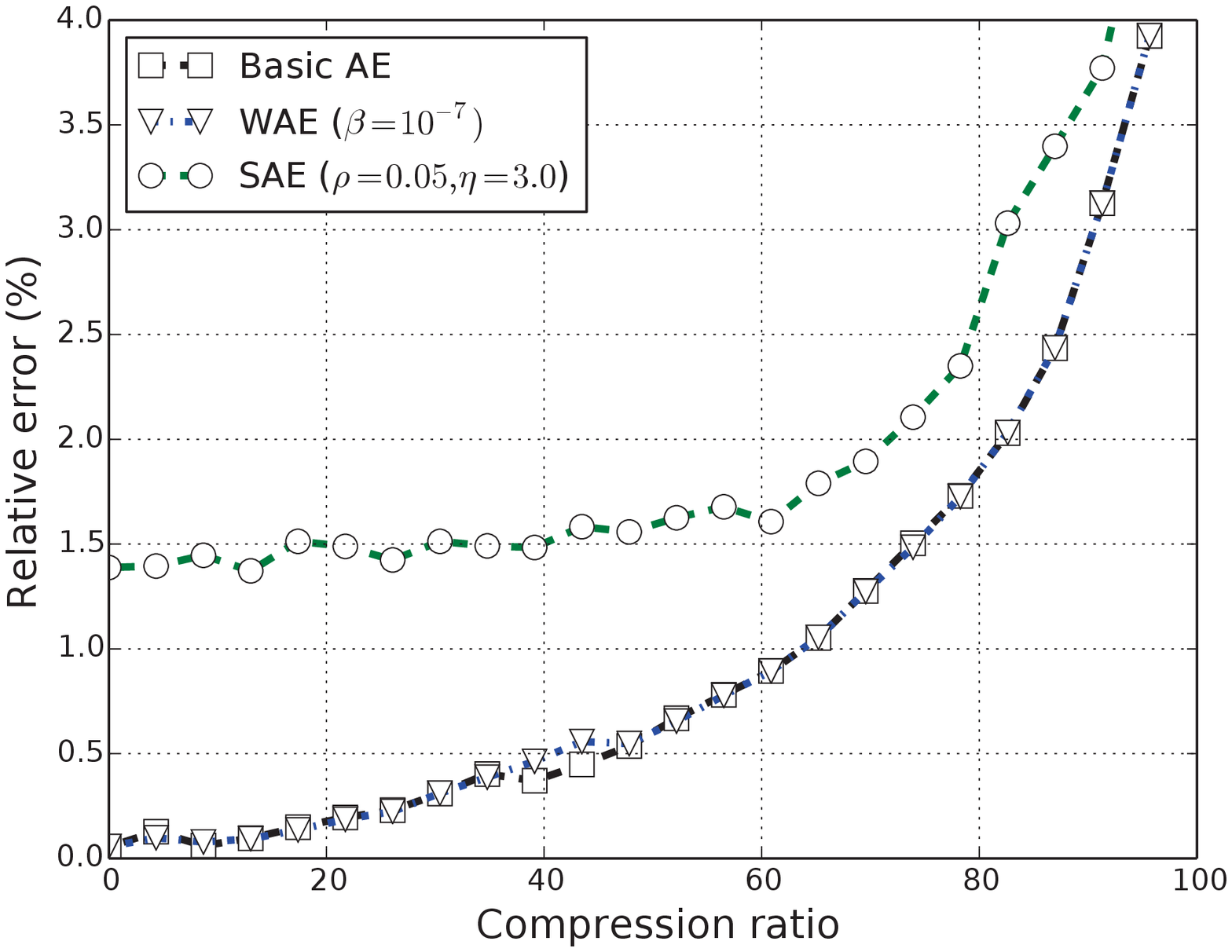}
\par\end{centering}

}\enskip{}\subfloat[{\label{fig:experimental_validation_stbernard_2}\small{}Spatial compression: Compression ratio vs. mean absolute error $\epsilon_{abs}$ (logarithmically spaced).}]{\begin{centering}
\includegraphics[width=0.33\columnwidth]{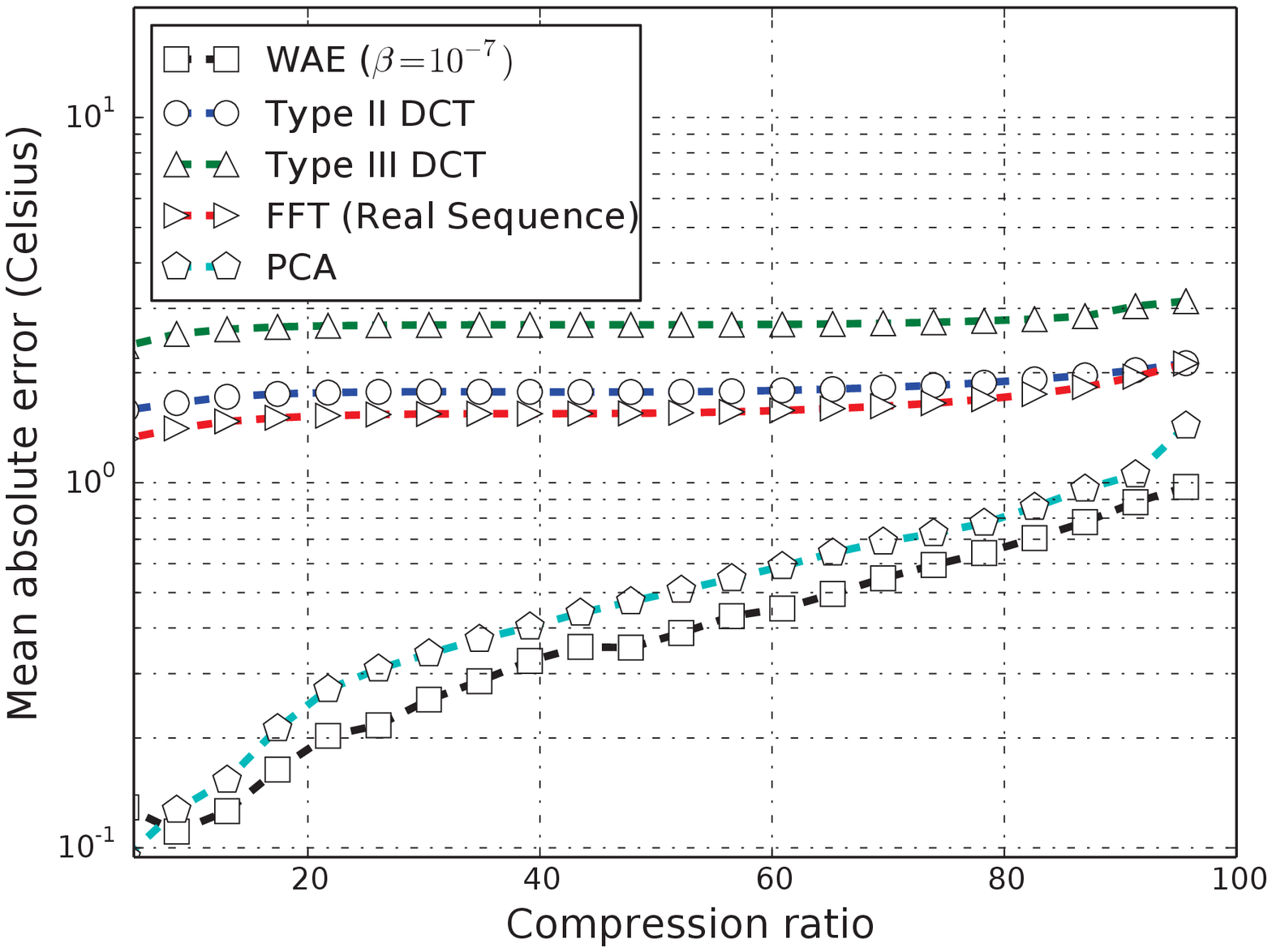}
\par\end{centering}

}\enskip{}\subfloat[{\label{fig:experimental_validation_luce_2}\small{}Temporal compression: Error bound $\epsilon_{bound}$ vs. compression ratio}]
{\begin{centering}
\includegraphics[width=0.33\columnwidth]{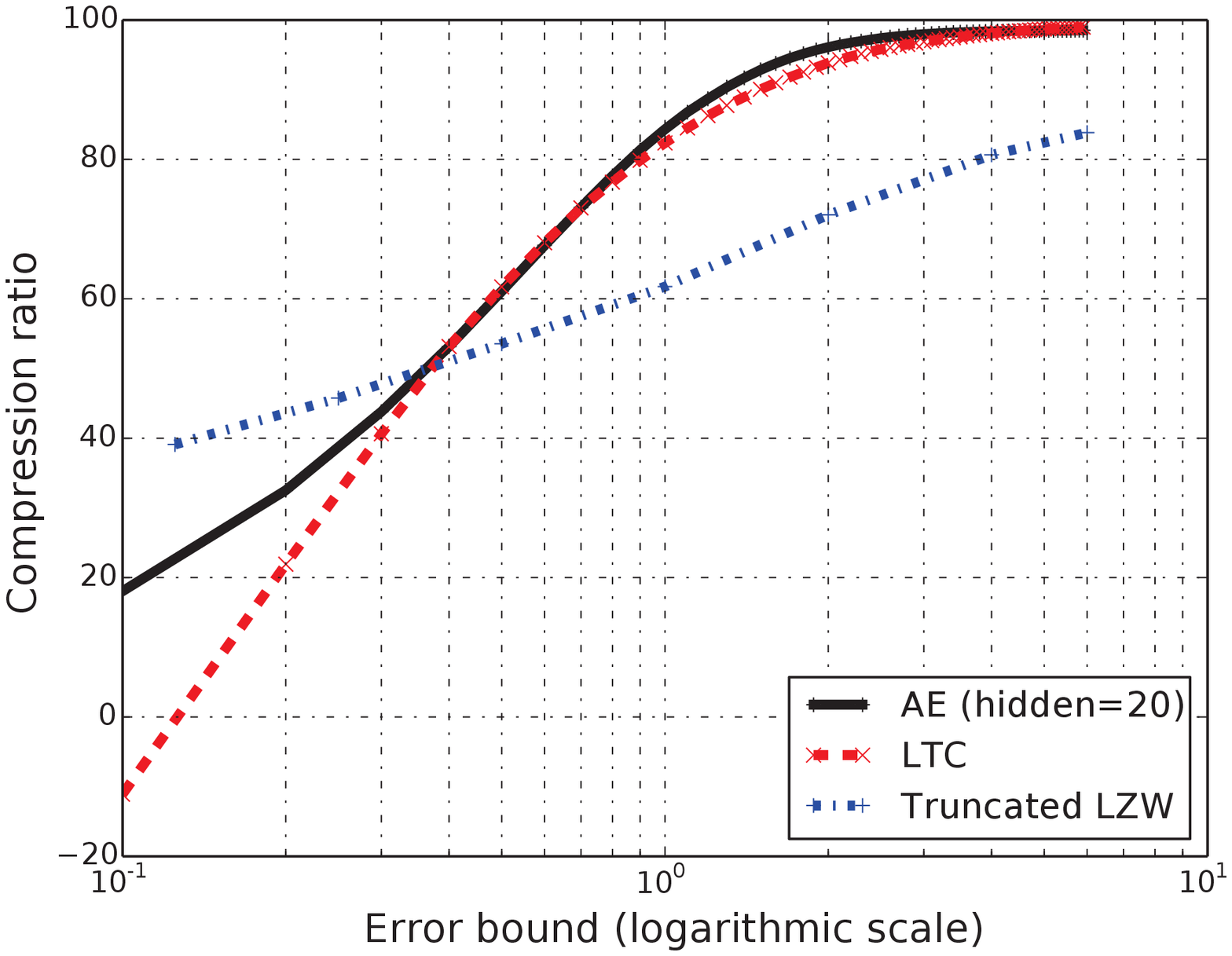}
\par\end{centering}

}
\par\end{centering}
\vspace{-3mm}

\caption{\label{fig:experimental_validation_stbernard}{\small{}Experimental results and validation of the spatial and temporal compression techniques.}}
\end{figure*}

\subsection{Overall performance}

As shown in Figure \ref{fig:experimental_validation_stbernard}, our algorithm demonstrates better performance on real world data sets when compared to traditional methods for data compression in WSNs. The data compression is considered as a challenging task due to the non-uniform data distribution through different sensor nodes. Comparatively, using basic AE or WAE provides the best performance over the other AE's variants (Figure \ref{fig:experimental_validation_stbernard_1}). Moreover, WAE outperforms other traditional compression methods such as PCA, DCT and FFT (Figure \ref{fig:experimental_validation_stbernard_2}).

LZW is commonly used as a basis for comparison against other data compression algorithms. In our modified method, we first convert the base-10 floating point readings into the base-2 representation, e.g., 10.51 is represented as 00001010.1 under 0.1 error bound. As a result, the truncated LZW algorithm can be realized as a lossy data compression scheme with a compression ratio that significantly outperforms the traditional LZW method. Moreover, we chose LTC algorithm for bench-marking as several comparative studies, e.g., \cite{zordan2012compress}, discussed the efficiency of the LTC algorithm over other methods. Even though the used high resolution data set is very suitable for the LTC method as the data changes slowly between subsequent samples, the compression efficiency of the proposed algorithm is still superior (Figure \ref{fig:experimental_validation_luce_2}). We note that LTC performs as well as AE for large error bounds, but is unable to keep up when the error bound is small. On the other hand, the truncated LZW does well for small error bounds since it is suited for lossless compression, but fails to handle large error bounds. Moreover, the truncated LZW is more computationally- and memory-intensive than AE, making it unsuitable for simple sensor nodes.

\section{Conclusion}\label{sec:conclusion}
Instead of using computationally expensive transformations on raw data or introducing strong assumptions on data statistical models, we proposed an adaptive data compression with feature extraction technique using AEs. Our solution exploits spatial-temporal correlations in the training data to generate a low dimensional representation of the raw data, thus significantly prolonging the lifespan of data aggregation and funneling systems. Moreover, the algorithm can optionally be adjusted to support error bound guarantee.

\bibliographystyle{abbrv}
\balance
\bibliography{short_mswim}

\end{document}